\documentclass[]{aa}
\usepackage{graphicx}
\begin{document}
\title{CCD photometric search for peculiar stars in open clusters. VII. 
Berkeley~11, Berkeley~94, Haffner~15, Lyng{\aa}~1, NGC~6031, NGC~6405, NGC~6834 and Ruprecht~130 
\thanks{Based on 
observations at the BNAO Rozhen, CASLEO, CTIO (Proposal
2003A-0057), ESO-La Silla (Proposal 073.C-0144)}}
\author{E.~Paunzen\inst{1}, M.~Netopil\inst{1}, I.Kh.~Iliev\inst{2}, H.M.~Maitzen\inst{1}, 
A.~Claret\inst{3}, O.I.~Pintado\inst{4,}
\thanks{Visiting Astronomer at Complejo Astron\'omico El Leoncito operated under agreement between 
Consejo Nacional de Investigaciones Cient\'ificas y T\'ecnicas de la Rep\'ublica Argentina and the 
National Universities of La Plata, Cordoba and San Juan.}}

\mail{Ernst.Paunzen@univie.ac.at}

\institute{Institut f\"ur Astronomie der Universit\"at Wien,
           T\"urkenschanzstr. 17, A-1180 Wien, Austria
\and	   Institute of Astronomy, National Astronomical Observatory, 
           P.O. Box 136, BG-4700 Smolyan, Bulgaria
\and	   Instituto de Astrof\'isica de Andaluc\'ia
		   CSIC, Apartado 3004, 18080 Granada, Spain
\and	   Departamento de F\'isica, Facultad de Ciencias Exactas 
           y Tecnolog\'ia, Universidad Nacional de Tucum\'an, Argentina - Consejo Nacional 
		   de Investigaciones Cient\'ificas y T\'ecnicas de la Rep\'ublica Argentina}

\date{Received 02 December 2005 / Accepted 05 February 2006}
\authorrunning{E. Paunzen et al.}{}
\titlerunning{Photometric search for peculiar stars in open clusters. VII.}{}

\abstract{}
{The detection of magnetic chemically peculiar (CP2) stars in open
clusters of the Milky Way can be used to study the influence of different galactic
environments on the (non-)presence of peculiarities, which has to be taken into
account in stellar evolution models. Furthermore it is still unknown if the
CP2 phenomenon evolves, i.e. does the strength of the peculiarity 
feature at 5200\AA\, increase or decrease with age.}
{We have observed eight young to intermediate age open
clusters in the $\Delta a$ photometric system. This intermediate 
band photometric system 
samples the depth of the 5200\AA\, flux depression by comparing the flux at the center
with the adjacent regions having bandwidths of 110\AA\, to 230\AA.  
The $\Delta a$ photometric system is most suitable to detect CP2 stars with high
efficiency, but is also capable of detecting a small percentage of 
non-magnetic CP objects. Also, the groups of (metal-weak) $\lambda$ Bootis, 
as well as classical Be/shell stars, can be successfully investigated.
This photometric system allows one to determine the age, reddening and
distance modulus by fitting isochrones.}
{Among the presented sample of eight galactic clusters, we have detected 
twenty three CP2, eight Be/Ae and eight metal-weak stars. Another six objects 
show a peculiar behaviour which is most probably due to a non-membership,
variability or duplicity. Fitting isochrones to $\Delta a$ photometry yields
estimates of the age, reddening and distance that are in excellent agreement
with published values.}
{}


\maketitle

\section{Introduction}
The continuation of our CCD $\Delta a$ photometric survey to detect 
chemically peculiar (CP) stars of the upper main sequence currently comprises the largest sample 
of open clusters, including young and intermediate age clusters at a variety 
of galactic longitudes and galactocentric distances. 

The $\Delta a$ photometric system allows one to efficiently detect CP stars and related objects 
through the flux depression at 5200\,\AA$ $ (Kupka et al. 2004). 
The efficiency of this system was recently investigated by Paunzen et al. (2005a) surveying a large 
sample of published field star measurements, resulting in a probability of up to 95\,\% of 
detecting all relevant magnetic CP stars. 
Furthermore the groups of $\lambda$Bootis and classical Be/shell systematically show negative $\Delta a$ 
values.

Unambiguous detection or non-detection of CP stars in different galactic environments 
will help to understand their evolution and formation. The aim of this survey is to investigate 
the occurence of CP stars in open clusters depending on various parameters like age, metallicity 
and galactic location. Preliminary results have shown that the 
percentage of CP2 stars with ages between 30 and 100\,Myr have two clear maxima, separated by a deep 
minimum. Therefore few of open clusters in this paper were selected within that age range 
to clarify this situation.

Beside the detection of CP stars, we are able to determine cluster parameters like the age, 
reddening and distance using isochrones for the $\Delta a$ system (Claret et al. 2003, 
Claret 2004) which were compared with already published parameters, yielding excellent agreement.

In the presented sample of eight galactic clusters (Berkeley~11, Berkeley~94, 
Haffner~15, Lyng\aa~1, NGC~6031, NGC~6405, NGC~6834 and Ruprecht~130) we have detected 
twenty three CP2, eight Be/Ae, eight metal-weak and six objects showing peculiar 
behaviour due to doubtful membership, variability or binarity. 

\section{Observations, reduction and used methods}

Observations of the eight open clusters were performed 
at four different sites and telescopes: 
\begin{itemize}
\item 2\,m RC telescope (BNAO, Rozhen),
direct imaging, SITe SI003AB 1024\,$\times$\,1024 pixel CCD and VersArray 1300B camera with EEV chip
1340\,$\times$\,1300 pixels,
5$\arcmin$ field-of-view
\item 0.9\,m telescope (CTIO), direct imaging, SITe 2084\,$\times$\,2046 pixel CCD,  
13$\arcmin$ field-of-view
\item 3.6\,m telescope (ESO-La Silla), EFOSC2, Loral/Lesser 2048\,$\times$\,2048 pixel
CCD, 5$\arcmin$ field-of-view
\item 2.15\,m RC telescope (CASLEO), direct imaging, EEV CCD36-40 1340\,$\times$\,1300 pixel 
CCD, 9$\arcmin$ field-of-view
\end{itemize}
The observing log with the number of frames in each filter
is listed in Table \ref{log}. The observations were performed with
two different filter sets, both having the following characteristics:
$g_1$ ($\lambda_c$\,=\,5007\,\AA, FWHM\,=\,126\,\AA, $T_P$\,=\,78\%), 
$g_2$ (5199, 95, 68) and $g_3$\,=\,$y$ (5466, 108, 70).

The basic CCD reductions and a point-spread-function fitting 
were carried out within standard IRAF V2.12.2 routines on
Personal Computers running under LINUX.
The way calculation of the normality line, derivation of the errors
as well as the calibration of our $(g_1-y)$ as well as $y$ measurements 
is the same as in previous works (see Netopil et al. 2005 and Paunzen et al. 2005b).  
 
\begin{table}[t]
\begin{center}
\caption{Observing log for the programme clusters. All clusters were observed on one night by 
O.I.~Pintado (OIP), M.~Netopil (MN), H.M.~Maitzen (HMM) and I.Kh.~Iliev (IKI).}
\label{log}
\begin{tabular}{lcccccc}
\hline\hline
Cluster & Site & Date & Obs. & \#$_{g_{\rm 1}}$ & \#$_{g_{\rm 2}}$ & \#$_{y}$ \\
\hline
Be~11 & BNAO & 02.2005 & IKI & 10 & 10 & 10 \\
Be~94 & BNAO & 08.2004 & IKI & 8 & 8 & 8 \\
Ha~15 & CTIO & 04.2003 & HMM & 11 & 11 & 12 \\
Ly~1 & ESO & 06.2004 & MN & 10 & 10 & 10 \\
N~6031 & ESO & 06.2004 & MN & 10 & 10 & 10 \\
N~6405 & CTIO & 04.2003 & HMM & 8 & 8 & 8 \\
N~6834 & BNAO & 08.2004 & IKI & 13 & 14 & 14 \\
Ru~130 & CASLEO & 06.2003 & OIP & 10 & 10 & 10 \\
\hline
\end{tabular}
\end{center}
\end{table}

The isochrones shown in Figs. \ref{fig1} to \ref{fig3} were taken from
Claret et al. (2003) and Claret (2004) and are based on the
$\Delta a$ photometric system. The derived ages, reddening and
distance moduli together with the errors are listed in Table \ref{all_res}.
The fitting procedure takes advantage of the available $UBV$ measurements for
all programme clusters which means that the results were compared with
those of the color-magnitude-diagrams for the $UBV$ photometric system.
However, our determination is based on the $\Delta a$ measurements
alone, which is another important application of this photometric system.

\begin{table*}[t]
\begin{center}
\caption{Summary of results; the age, distance modulus, reddening
and thus the distance from the Sun was derived by fitting isochrones
to the $\Delta a$ photometry. $R_{V}$ was set to 3.1. For the distance of the Sun from
the galactic center, a value of 8.5\,kpc was used, the Trumpler classification was taken 
from Lyng{\aa} (1987) and the errors in the final digits of the corresponding quantity
are given in parenthesis. Questionable peculiar stars are marked by asterisks (see the 
corresponding cluster result section for details). All photometric values are in units of mmags.}
{\scriptsize
\label{all_res}
\begin{tabular}{lllll}
\hline\hline
Name & Berkeley~11 & Berkeley~94 & Haffner~15 & Lyng{\aa}~1 \\
     & \object{C0417+448} & \object{C2220+556} & \object{C0743$-$326} & \object{C1356$-$619} \\
\hline
$l/b$ & 157.08/$-$3.64 & 103.13/$-$1.18 & 247.94/$-$4.16 & 310.86/$-$0.35\\
$E(B-V)$ ($\pm$0.05) & 0.95 & 0.65 & 1.10 & 0.50 \\
$m_V - M_V$ ($\pm$0.2) & 15.2 & 14.6 & 15.1 & 13.0 \\
$d$\,[kpc] & 2.82(46) & 3.29(54) & 2.21(36) & 1.95(32) \\
$R_{GC}$\,[kpc] & 11.15(44) & 9.79(30) & 9.55(21) & 7.37(15) \\
$|z|$\,[pc] & 179(29) & 68(11) & 160(26) & 12(2) \\
log\,$t$ ($\pm$0.1) & 8.0 & 7.0 & 7.2 & 8.0  \\
Tr-type & II 2 m & II 3 p & II 2 m & II 2 p\\
n(member) & 54 & 77 & 83 & 40 \\
n(none) & 12 & 38 & 246 & 126 \\   
CP No. (Webda):  & *25(322): +91/+123/+911      & 27($-$): $-$54/$-$/+2737  & 109(3): +45/$-$214/$-$3078  & 8(58): +48/+371/+3153  \\
                                $\Delta$a/$(B-V)_0$/$M_V$  & 38(188): +45/$-$84/$-$631   & 67(4): +42/$-$/+696    & 190($-$): +65/$-$129/1867     &  15(53): +26/+280/+2880 \\
                                 $[$mmag$]$ & 50(128): +42/$-$146/$-$2000 & 114($-$): +88/$-$/+1719    &                    & 22(56): $-$44/+498/+3016\\
                                  & *54(255): +111/+109/+384     &               &                    &  81(80): $-$25/+448/+3710 \\
                                  & 64(313): +33/$-$130/$-$1390  &              &                    & *82(40): +51/+187/+1510  \\
                                  &                     &              &                    & 93(57): +22/+307/+3132  \\
                                  &                     &              &                    & 103(52): +37/+253/+2801  \\
                                  &                     &              &                    & 105(24): $-$35/+130/+1089  \\
                                  &                     &              &                    & 109(125): $-$23/+512/+4570  \\
                                  &                     &              &                    & 111(87): $-$21/+523/+3909  \\
                                  &                     &              &                    & 112(85): +22/+388/+3912 \\
                                  &                     &              &                    & 127(31): $-$43/+143/+1556  \\
                                  &                     &              &                    & 134(67): $-$41/+420/+3378  \\
n(frames) & 30  & 24  & 34  & 30  \\

\hline\hline
Name & NGC~6031 & NGC~6405 & NGC~6834 & Ruprecht~130 \\
     & \object{C1603$-$539} & \object{C1736$-$321} & \object{C1950+292} & \object{C1744$-$301} \\
\hline
$l/b$ & 329.29/$-$1.55 & 356.62/$-$0.74 & 65.71/+1.19 & 359.21/$-$0.98\\
$E(B-V)$ ($\pm$0.05) & 0.50 & 0.20  & 0.70 & 1.20 \\
$m_V - M_V$ ($\pm$0.2) & 12.45  & 8.9 & 13.6 & 15.0  \\
$d$\,[kpc] & 1.51(25) & 0.45(7) & 1.93(32) & 1.80(30)  \\
$R_{GC}$\,[kpc] & 7.24(20) & 8.05(7) & 7.90(6) & 6.70(30) \\
$|z|$\,[pc] & 41(7) & 6(1) & 40(7) & 31(5) \\
log\,$t$ ($\pm$0.1) & 8.2 & 8.0 & 7.9 & 7.9  \\
Tr-type & I 3 p & II 3 r & II 2 m & II 1 p \\
n(member) & 67 & 134 & 201 & 140 \\
n(none) & 176 & 33 & 107 & 155 \\
CP No. (Webda):      & 29(85): +28/+151/+2651  & *33(7): +32/+57/+1915  & 8(128): $-$82/+30/+1854  & 71($-$): +35/+83/$-$285  \\	
$\Delta$a/$(B-V)_0$/$M_V$    & 35(73): +20/+24/+1034    & 65(19): +55/$-$90/+839  & *32(125): $-$42/+771/$-$2637 & 79(1): +33/$-$187/$-$169   \\	
$[$mmag$]$                         & 69(70): $-$29/+387/+3017    & 135(77): +85/$-$94/+357  & *129(1): $-$46/+196/$-$3158  & 92(7): +48/$-$102/$-$518  \\	
                             & 199(54): $-$23/+583/+2893  &      & 137($-$): +41/$-$64/+1706  & 227($-$): +48/$-$72/+1157   \\	
                             &                            &      & 139(412): +36/$-$31/+2797  & 250(108): +60/$-$345/$-$2997  \\	
                             &                            &      & 198(67): $-$78/+97/$-$1128 &             \\	
                             &                            &      & 202(42): $-$59/$-$68/+526 &             \\	
                             &                            &      & 234(55): $-$57/$-$34/$-$385 &             \\	
                             &                            &      & 294(155): $-$71/$-$124/+1739 &             \\	
                             &                            &      & 297(73): $-$72/$-$35/$-$1963 &             \\	
n(frames) & 30  & 24  & 41  & 30  \\
\hline
\end{tabular}
}
\end{center}
\end{table*}

The tables with all data for the individual cluster stars as well as
nonmembers are available
in electronic form at the CDS via anonymous ftp to cdsarc.u-strasbg.fr (130.79.128.5),
http://cdsweb.u-strasbg.fr/Abstract.html, at WEBDA (http://www.univie.ac.at/webda/)
or upon request from the first author. These tables include the cross
identification of objects from the literature, the $X$ and $Y$
coordinates within our frames, the observed $(g_{1}-y)$ and
$a$ values with their corresponding errors, $V$ magnitudes,
the $(B-V)$ colors from the literature,
$\Delta a$-values derived from the normality lines of $(g_{1}-y)$,
(disregarding non-members) and the number of observations, respectively.

The diagnostic diagrams for
all eight open clusters are shown in Figs. \ref{fig1} to \ref{fig3}. 
The normality lines and the confidence intervals corresponding to 99.9\,\%
are plotted. The detected peculiar objects are marked by
asterisks. Only members (filled circles) have been used to derive the
normality lines. The selection of these objects was done according to their
location in the color-magnitude diagrams as well as the distance from the
cluster centers and additional information
from the literature (proper motions and radial velocities) taken from
WEBDA.

\section{Results} \label{results}

\begin{figure*}
\begin{center}
\includegraphics[width=170mm]{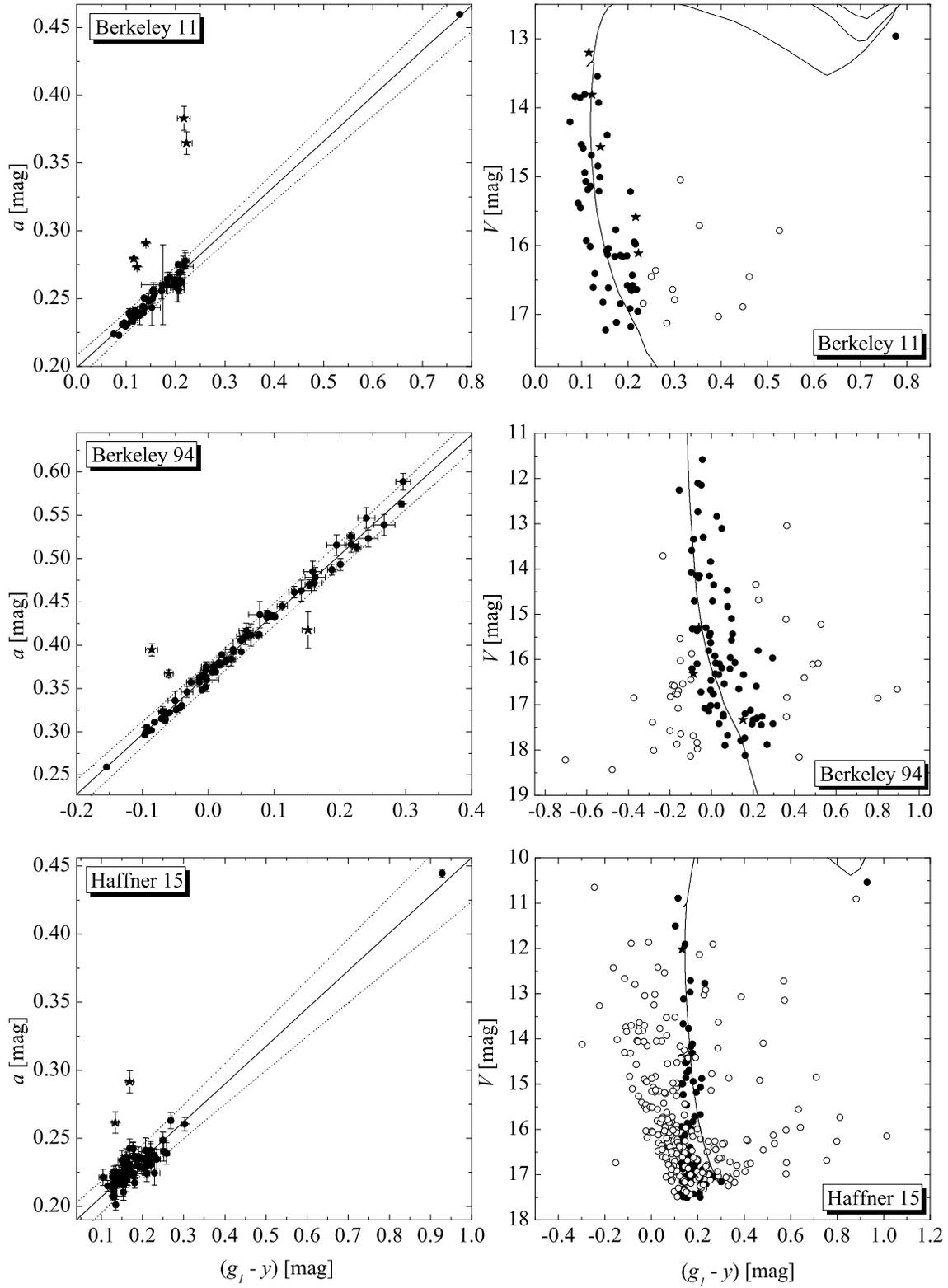}
\caption[]{Observed $a$ versus $(g_1-y)$ and $V$ versus $(g_1-y)$
diagrams for our programme clusters. The solid line is the
normality line whereas the dotted lines are the confidence intervals
corresponding to 99.9\,\% (left panels). The error bars for each individual object
are the mean errors. The detected peculiar objects are marked with
asterisks. Only members (filled circles) have been used to derive the
normality lines. The fitting parameters are listed
in Table \ref{coeffs}. The isochrones (right panels) were
taken from Claret et al. (2003) and Claret (2004) and are based on the
$\Delta a$ photometric system. The derived ages, reddening and
distance moduli are given in Table \ref{all_res}.}
\label{fig1}
\end{center}
\end{figure*}

\begin{figure*}
\begin{center}
\includegraphics[width=170mm]{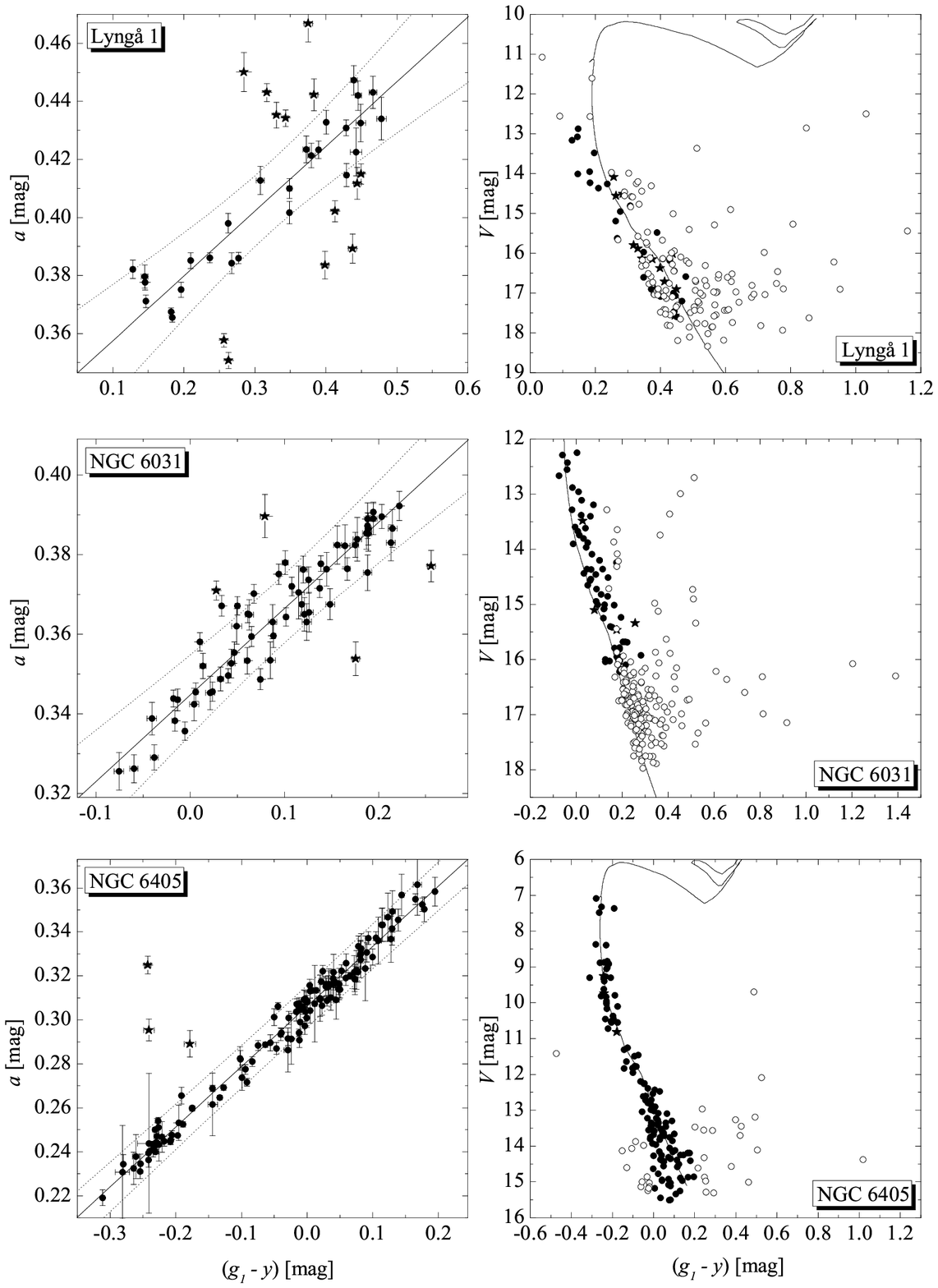}
\caption[]{Observed $a$ versus $(g_1-y)$ and $V$ versus $(g_1-y)$
diagrams for our programme clusters. The symbols
are the same as in Fig. \ref{fig1}.}
\label{fig2}
\end{center}
\end{figure*}

\begin{figure*}
\begin{center}
\includegraphics[width=170mm]{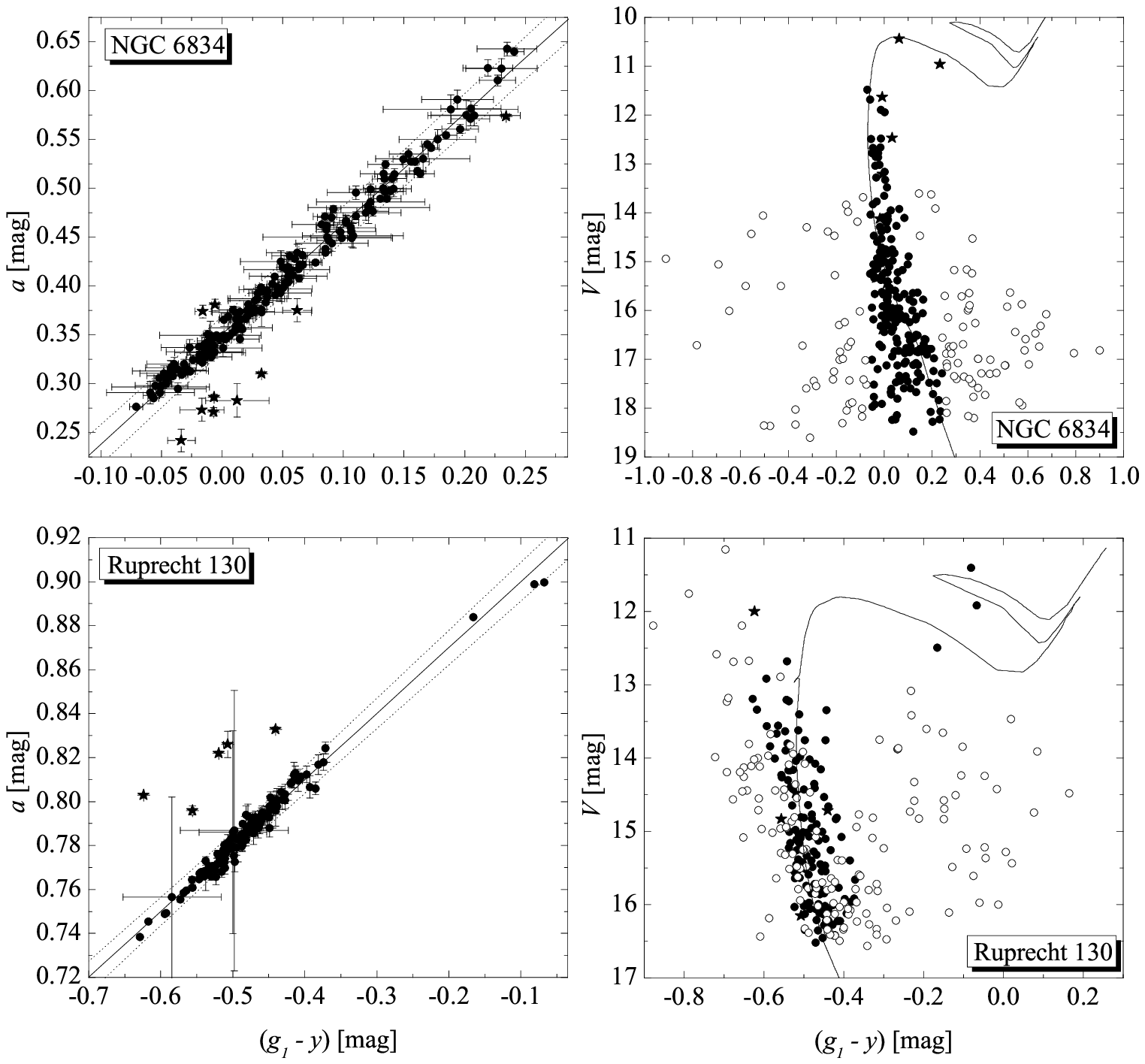}
\caption[]{Observed $a$ versus $(g_1-y)$ and $V$ versus $(g_1-y)$
diagrams for our programme clusters. The symbols
are the same as in Fig. \ref{fig1}.}
\label{fig3}
\end{center}
\end{figure*}

\begin{table}[t]
\begin{center}
\caption{Comparison of our results (bold face) with already published cluster 
parameters. The errors of the last digit are given in parentheses if available.  
If interstellar extinction was given in other photometric systems (* $RGU$, ** $uvby$) they 
were transformed to $E(B-V)$ as described in Section \ref{results}. The following references, 
sorted by cluster and year, are given in the last column: (1) Jackson, 
Fitzgerald \& Moffat (1980); (2) Yadav \& Sagar (2002); (3) Yilmaz (1970); (4) 
Wramdemark (1978); (5) Fitzsimmons (1993); (6) Vogt \& Moffat (1972); (7) 
Peterson \& Fitzgerald (1988); (8) V\'azquez et al. (2003); (9) Kharchenko et al. (2005); 
(10) Lindoff (1967); (11) Moffat \& Vogt (1975); (12) Fenkart \& Binggeli (1979); 
(13) Topatkas (1981); (14) Piatti et al. (1999); (15) Rohlfs et al. (1959); 
(16) Eggen (1961); (17) Talbert (1965); (18) Vleeming (1974); (19) Schneider (1985); 
(20) Johnson (1961); (21) F\"unfschilling (1967); (22) Moffat (1972); (23) Piatti et al. (2000)
}
\label{comparison}
\begin{tabular}{llllr}
\hline\hline
Cluster & log\,$t$ & $E(B-V)$ & $d$\,[kpc] & Ref \\
\hline
\textbf{Be~11}  & \textbf{8.0}(1) & \textbf{0.95}(5) & \textbf{2.82}(46) & \\
       &  7.5 & 0.96(6)  & 2.2(2) & 1\\
       &  8.05(5) & 0.95(5) & 2.2(1)  & 2 \\
\textbf{Be~94}  & \textbf{7.0}(1) & \textbf{0.65}(5) & \textbf{3.29}(54) & \\
       &   & 0.48*  & 1.61  & 3\\
       &  7.0 & 0.65  & 5  & 4\\
       &  6.8 & 0.65**  & 3.5  & 5\\
\textbf{Ha~15}  & \textbf{7.2}(1) & \textbf{1.10}(5) & \textbf{2.21}(36) & \\
       &   & 1.16(8)  & 2.49  & 6\\
\textbf{Ly~1}   & \textbf{8.0}(1) & \textbf{0.50}(5) & \textbf{1.95}(32) & \\
       & 7.9 & 0.45(2)  & 1.98  & 7\\
       & 8.05 &  0.45(3) & 1.9(1)  & 8\\
       & 8.15 & 0.46 & 2.28  & 9\\
\textbf{N~6031} & \textbf{8.2}(1) & \textbf{0.50}(5) & \textbf{1.51}(25) & \\
       & 7.5  &  0.43(3) & 3.2  & 10\\
       &   & 0.47(5)  & 1.09  & 11\\
       &   & 0.45  & 1.78  & 12\\
       &   & 0.33* & 1.59  & 13\\
       &  8.3(3) & 0.40(15) & 2.1(7)  & 14\\
\textbf{N~6405} & \textbf{8.0}(1) & \textbf{0.20}(5) & \textbf{0.45}(7) & \\
       & 8.0  & 0.16  & 0.63(5)  & 15\\
       &   & 0.13  & 0.50  & 16\\
       &   & 0.16  & 0.48(5)  & 17\\
       & 8.0  & 0.15 & 0.45(5) & 18\\
       &   & 0.15**  &   & 19\\
       & 7.91  & 0.14  & 0.49  & 9\\
\textbf{N~6834} & \textbf{7.9}(1) & \textbf{0.70}(5) & \textbf{1.93}(32) & \\
       &   & 0.72  & 3.03 & 20 \\
       &   & 0.61  & 2.1 & 21 \\
       & 7.9  & 0.72  & 2.14 & 22 \\
\textbf{Ru~130} & \textbf{7.9} & \textbf{1.20}(5) & \textbf{1.8}(30) &  \\
       & 7.7(1) & 1.20(5) & 2.1(4) &  23\\
\hline
\end{tabular}
\end{center}
\end{table}

In the following we will present the results and the comparison with
literature values for the individual open clusters. The 
cluster parameters are given in Table \ref{comparison}. 
Significant deviating cases are discussed in more detail. The errors of each quantity are 
given in Table \ref{all_res} and \ref{comparison} and will be not repeated here. 
For Berkeley~94, NGC~6031 and NGC~6405, estimates of interstellar reddening 
were published in the Becker $RGU$ and Str{\"o}mgren $uvby\beta$ system. These
values were transformed to $E(B-V)$ according to the following relations: 
$E(G-R)$\,=\,1.39$E(B-V)$, Steinlin (1968), and $E(b-y)$\,=\,0.74$E(B-V)$, Crawford (1978). 

The star numbers are from our numbering system based on ascending (X,Y) coordinates
on the CCD frames. The 
$\Delta a$ indices in units of mmag are always given in parenthesis. 
Table \ref{results} lists also 
the star numbers according to WEBDA, if available. 
Table \ref{coeffs} contains the slopes of the normality lines for the individual open clusters. From the 
theoretical work of Kupka et al. (2003) this slope ($a$ vs. $(g_{1}-y)$) should be $\sim$ 0.20 for the 
hotter stars. This is only moderately lower than most of our cluster values, with two major exceptions: 
Berkeley~94 and NGC~6834. Further observations are needed to clarify these substantial deviations.
\\

{\bf Berkeley~11:}
Five positive CP detections were obtained. The stars \#38 
(+45\,mmag), 50 (+42) and 64 (+33) are indicated as members in both previous studies of 
that cluster (Jackson et al. 1980, Yadav \& Sagar 2002), although the other two  
deviating objects \#25 (+91) and 54 (+111) are questionable also due to their extremely 
high $\Delta a$ indices. Both stars are shifted slightly to the red in the 
color-magnitude diagram (Figure \ref{fig1}), but were determined to be non-members by 
Yadav \& Sagar (2002), although \#54 is not significant outlying in the $(B-V)$
versus $(U-B)$ 
diagram. There are several large differences in the available 
photometry for that cluster. A comparison of the data by Yadav \& Sagar (2002) and 
Jackson et al. (1980), performed by Mermilliod \& Paunzen (2003), resulted in 
differences of $\Delta(B-V)$\,=\,$-$0.08 and $\Delta(U-B)$\,=\,$-$0.25, however the 
corresponding diagrams of the photometric comparison by Yadav \& Sagar (2002) 
do not show such a behaviour. The interstellar reddening obtained for the 
cluster, $E(B-V)$\,=\,0.95, is in agreement with all previous studies 
(Table \ref{comparison}). The determined age of log\,$t$\,=\,8.0 is comparable 
to Yadav \& Sagar (2002), whereas Jackson et al. (1980) classified the cluster as much 
younger. For the distance modulus we derived $m_V - M_V$\,=\,15.2, resulting in a distance 
of 2.82\,kpc which is slightly higher than in the literature (2.2\,kpc).
\\

{\bf Berkeley~94:}
the star \#67 (+42), determined as a physical cluster member by Yilmaz (1970), 
and one not previously measured star, \#114 (+88), could be identified as classical CP2 stars 
in the spectral range of about A5. \#27 ($-$54), seems to be a 
metal-weak star, since it is slightly too cool for a Be/Ae object. However, an estimation of 
their spectral type on the basis of their calibrated $(B-V)$ colors is difficult due to the 
small number and range of corresponding measurements in other photometric systems. None of 
the peculiar objects deviate from the cluster main sequence (Figure \ref{fig1}), 
which supports their membership status, but additional investigations are necessary. 
The determined age, log\,$t$\,=\,7.0, and interstellar reddening, 
$E(B-V)$\,=\,0.65, for Berkeley~94 are in agreement with the results from the literature. 
A comparison 
of the distances is more difficult. The distance of 3.3\,kpc ($m_V - M_V$\,=\,14.6) 
is only comparable to the value derived by Fitzsimmons (1993, 3.5\,kpc). 
The other two works (Yilmaz 1970; Wramdemark 1978) list 1.61 and 5\,kpc, respectively. 
These discrepancies deserve further attention.
Yilmaz (1970) stated that his published photometric measurements are rather uncertain due to
technical problems. Wramdemark (1978), on the other hand, analysed only
13 stars resulting in a poor representation of the main sequence.
This might cause the widely different distance estimations.
\\

{\bf Haffner~15:}
Two deviating stars were found within the poorly populated and investigated cluster 
Haffner~15. The objects \#109 (+45), a member according to Vogt \& Moffat (1972), 
and \#190 (+66) are candidates for classical CP2, mid B-type stars. Our obtained 
values for interstellar reddening, $E(B-V)$\,=\,1.10, and distance, 2.2\,kpc 
($m_V - M_V$\,=\,15.13), 
are in agreement with the result Vogt \& Moffat (1972). The age was determined for the first time to our 
knowledge as log\,$t$\,=\,7.2, however the catalogue by Dias et al. (2002) 
lists an almost identical age of log\,$t$\,=\,7.17 for that cluster.
The obtained CCD-photometry displayed in Fig. \ref{fig1} shows stars in the cluster area within
a diameter of 3$\arcmin$ by filled circles. Open circles denote objects
which by their position in the HR-diagram should be considered as non-members
of Haffner~15. Since we notice a blue branch among them, suggesting the
presence of an even younger stellar aggregate in the vicinity of Haffner~15, we 
investigate this possibility by exploring its local relationship to the cluster.
We found that these ``young'' stars are, indeed, not distributed equally over
the image area, but are concentrated in the southwest of the cluster. On the other hand,
a stellar group significantly younger than Haffner~15 would imply 
star formation, hence interstellar material with an H$\alpha$ signature.
Inspection of the POSS data excluded this possibility.
From the distribution of interstellar extinction (Neckel \& Klare 1980) we
deduce from Field 94 (centered on $l$\,=\,247$\degr$ and $b$\,=\,$-$5$\degr$) that 
beyond the distance of
1000\,pc the field exhibits extreme diversity: the values cover the
interval from less than 1 to nearly 4 magnitudes of visual absorption.
We conclude that the allegedly young stars,
despite of their angular coherence, are not an expanding (young)
association, but are located along the reddening vector direction in Fig. \ref{fig1} 
with lower interstellar extinction than the cluster, thus mimicking a
stellar aggregate younger than Haffner~15.
\\

{\bf Lyng{\aa}~1:} was recently investigated by V{\'a}zquez et al. 
(2003) who reported the finding of a red supergiant member, but unfortunately we are not able to 
measure that star due to saturation. Since it is difficult to fit an appropriate isochrone 
without it, we have used log\,$t$\,=\,8.0, roughly the average of previous studies, to 
fit the remaining parameters and determined them as $E(B-V)$\,=\,0.50 and $m_V - M_V$\,=\,13.0 
which is consistent with the literature values. Since about two thirds of the probable 
members according to V{\'a}zquez et al. (2003) exhibit peculiar behaviour, we 
set a cut off at late F-type based on calibrated $(B-V)$ color indices 
and the obtained reddening value. This is justifiable because CP stars are only expected up to a 
spectral type of F5. Using this restriction, thirteen stars (see Table \ref{all_res} for the 
complete listing) out of forty probable members deviate from the normality line, still 
an extremely high number. The objects \#105 ($-$35) and 127 ($-$43) are candidates on the 
border to be Ae stars, whereas the other five objects showing a negative $\Delta a$ index 
are possible metal-weak stars. The six detected CP2 stars are in the spectral range of 
about A6 to F4 with $\Delta a$ indices between +22 and +51\,mmag. One suspected variable 
star, \#82 (+51), was found within that group of peculiar objects. Our calibrated 
$V$ magnitude deviates compared to all available measurements: +0$\fm$36  
(Peterson \& Fitzgerald 1988) and $-$0$\fm$53 (V{\'a}zquez et al. 2003). This object is 
therefore marked as questionable in Table \ref{all_res}.  
Further membership and spectroscopic investigations for this cluster are planned
in the near future. 
\\

{\bf NGC~6031:}
Since a few stars lie outside the confidence interval, an $\Delta a$ index of 20\,mmag 
was set as the threshold for peculiarity. On the basis of that limit, four objects 
show a peculiar behaviour. The stars \#29 (+28) and 35 (+20) are candidates for classical 
CP2 stars, whereas the stars \# 69 ($-$29) and 199 ($-$23) are metal-weak objects, since they 
are too cool to be Be/Ae objects. All deviating stars are members according to Topatkas (1981) 
and/or Lindoff (1967). The objects fainter than about 16$\fm$2 seem to be field stars because 
they deviate extremely from the normality line. Such a field population
exhibits a main sequence in the color-magnitude diagram (Piatti et al. 1999). 
Our determined reddening of $E(B-V)$\,=\,0.50 is in agreement with previous results, 
whereas the other cluster 
parameters deviate between the different investigations. The age was formerly defined 
as log\,$t$\,=\,7.5 by Lindoff (1967) and 8.3 (Piatti et al. 1999) which correspond 
to our result of log\,$t$\,=\,8.2.  
For the cluster distance the result of 1.5\,kpc ($m_V - M_V$\,=\,12.45) is an average of 
the former studies, excluding Lindoff (1967) and Piatti et al. (1999) who obtained the  
larger distances of 3.2 and 2.1(7)\,kpc, respectively.     
\\

{\bf NGC~6405:}
The open cluster NGC~6405 was investigated by Maitzen \& Schneider (1984) who obtained 
photoelectric $\Delta a$ measurements. It was included in our CCD-survey in order to
compare the detection performance of both techniques employed, yielding excellent agreement of the three 
stars \# 33 (+32/+33), 65 (+55/+67) and 
135 (+85/+94) detected in both investigations. The $\Delta a$ indices in mmag of the present and previous 
study are given within brackets. A very good agreement is seen although there is
a decrease of the $\Delta a$-effect by about 10\% for at least two objects. This might be explained by
the intrinsic variability of these objects.

A more reasonable argument is the difference in the filter set used by
Maitzen \& Schneider (1984), who had their longest wavelength filter displaced by about 30\AA$ $ to the red, 
therefore better representing the continuum level.
Two stars (\#65 \& 135) were also recognized as CP2 stars within the 
Geneva photometric system (North \& Cramer, 1981). The remaining star \#33 seems to be 
a nonmember since it is an outlier in the two color $(B-V)$ versus $(U-B)$ diagram, but this object 
will be investigated spectroscopically in the near future to clarify its status. The 
cluster parameters determined via $\Delta a$ isochrone fitting, log\,$t$\,=\,8.0, 
$E(B-V)$\,=\,0.2 and $m_V - M_V$\,=\,8.9 are in concordance with all previous 
studies (see Table \ref{comparison}).
\\

{\bf NGC~6834:}
Our determined parameters, log\,$t$\,=\,7.9, $E(B-V)$\,=\,0.70 and $m_V - M_V$\,=\,13.6 are in 
good agreement with the results found in the literature, except Johnson (1961) who obtained 
a much larger distance. The age is identical to the one listed by Moffat (1972), 
the only available work that 
has investigated the cluster age. In total ten peculiar objects were found  
(see Table \ref{all_res}) among possible members according to their location within the 
color-magnitude diagram or the investigation by F\"unfschilling (1967). The evolved 
stars \#32 ($-$42) and 129 ($-$46), showing a possible emission, were investigated 
spectroscopically by Sowell (1987), classified as G0/5 III/V and F2 Ib, respectively. 
The object \#32 was determined as a possible foreground star and \#129 showed an increase 
in brightness in the $B$ and $V$ passbands of 0.2 to 0.3\,mag during a period of ninety 
days, however \#129 is a double star according to Mason et al. (2002). Both objects are 
therefore questionable and are marked in Table \ref{all_res} by asterisks. All other 
``negative'' stars are candidates for Be/Ae objects, whereas \#198 ($-$78), 234 ($-$57) and 297 
($-$72) can be also found in the catalogue of H$\alpha$ emission stars by 
Kohoutek \& Wehmeyer (1999). In addition, two positive CP2 detections for the 
objects \#137 (+41) \& 139 (+36) were observed.   
\\

{\bf Ruprecht~130:}
Five possible CP2 stars were detected in the cluster Ruprecht~130, \#71 (+35), 79 (+33), 
92 (+48), 227 (+48) and 250 (+60). Unfortunately no membership analysis is available in 
the literature. However, the detected CP2 stars are not outstanding within the color 
magnitude diagram (Fig. \ref{fig3}), except star \#250 which appears much bluer. This 
can be interpreted by the ``blueing'' effect, typical for some magnetic CP objects 
due to stronger UV absorption than in normal stars (Adelman 1980). For the determined 
cluster parameters, log\,$t$\,=\,7.9, $E(B-V)$\,=\,1.20, $m_V - M_V$\,=\,15.0, only 
marginal discrepancies from the study by Piatti et al. (2000) can be found. The 
isochrone fitting is difficult without knowledge of the membership status for the 
three evolved stars. However, their location close to the normality line supports 
membership.

\begin{table}[t]
\begin{center}
\caption{The regression coefficients for the
transformations and normality lines. The absolute values and errors
vary due to the inhomogeneous ``standard'' observations (photographic, photoelectric
and CCD) found in the literature as well as the dependence on the 
magnitude range in common, i.e. a broader range guarantees a small error. 
The offsets are due to the four different telescopes and thus instruments as well
as CCD used (Table \ref{log}), the deviating results for the slopes are discussed in 
the beginning of Section \ref{results}. 
The errors in the final digits of the corresponding quantity
are given in parentheses.}
\label{coeffs}
\begin{tabular}{ll|l}
\hline\hline
Cluster & $V$\,=\,a\,+\,b$\cdot(y)$, $N$ & $a_{0}$\,=\,a\,+\,b$\cdot(g_{1}-y)$, $N$ \\
\hline 
Be~11 & $-$3.54(13)/1.01(1)/66 & 0.199(1)/0.334(6)/49 \\
Be~94 & $-$2.70(36)/0.93(2)/14 & 0.366(1)/0.692(7)/74 \\
Ha~15 & $-$5.46(37)/1.01(2)/13 & 0.179(2)/0.277(9)/79 \\
Ly~1 &  $-$0.26(4)/1.01(1)/127 & 0.335(5)/0.223(14)/25 \\
N~6031 & $-$0.12(4)/1.01(1)/151 & 0.345(1)/0.217(9)/63 \\
N~6405 & $-$5.88(11)/1.03(1)/107 & 0.306(1)/0.274(3)/131 \\
N~6834 & $-$2.70(33)/0.95(2)/120 & 0.351(1)/1.129(9)/192 \\
Ru~130 & $-$5.32(6)/1.00(1)/111 & 0.930(2)/0.299(4)/136 \\
\hline
\end{tabular}
\end{center}
\end{table}

\section{Conclusions}

Within the presented sample of eight galactic open clusters, 1689 objects on 243 frames using 
four different sites were investigated, resulting in the detection of twenty three CP2, eight 
Be/Ae (three previously identified) and eight metal-weak objects. A further six deviating stars are 
designated as questionable either due to their membership status or because they have been identified as 
variable or double stars.   

As an important application of the $\Delta a$ photometric system, 
isochrones were fitted to the color-magnitude-diagrams ($V$ versus
$(g_1-y)$) of the programme clusters. For this purpose, our measured
$y$ magnitudes were directly converted into standard $V$ magnitudes
on the basis of already published values. A comparison of our results
yields an excellent agreement with the corresponding parameters from
the literature. 

The programme clusters yield a wide spread of different galactic environments (galactic 
longitude, galactocentric distances) and ages (10 to 160\,Myr), useful for the ongoing 
investigation on the incidence of peculiar stars. There is special interest in the cluster 
Lyng{\aa}~1, which has an extremely large number of peculiar stars with respect to the 
quantity of probable members ($\sim$ 30\%), but further spectroscopic and membership 
investigations are required to support this finding.

\begin{acknowledgements}
This research was performed within the projects  
{\sl P17580} and {\sl P17920} of the Austrian Fonds zur F{\"o}rderung der 
wissen\-schaft\-lichen
Forschung (FwF) and benefited also from financial contributions of the City of
Vienna (Hochschuljubil{\"a}umsstiftung projects: H-1123/2002). M.~Netopil acknowledges 
the support by a ``Forschungsstipendium'' from the University of Vienna and
I.Kh.~Iliev by the Bulgarian National Science Fund under grant F-1403/2004.
Use was made of the SIMBAD database, operated at the CDS, Strasbourg, France and
the WEBDA database, operated at the University
of Vienna. This research has made use of NASA's Astrophysics Data System.
\end{acknowledgements}

\end{document}